\documentclass[3p,times]{elsarticle}

\usepackage{ecrc}

\def\bs{\boldsymbol}

\def\p{{\boldsymbol p}_{\perp }}
\def\pb{\bar {\boldsymbol p}_{\perp }}

\def\q{{\boldsymbol q}_\perp}

\def\k{{\boldsymbol k}_\perp}

\def\x{{\boldsymbol x}_\perp}

\def\v{{\boldsymbol v}_\perp}

\def\bkappa{{\boldsymbol \kappa}_\perp}
\def\bbkappa{\bar{\boldsymbol \kappa}_\perp}

\def\qqb{{q\bar q}}
\def\sM{{\scriptscriptstyle M}}

\newcommand{\beq}{\begin{eqnarray}}
\newcommand{\eeq}{\end{eqnarray}}
\newcommand{\be}{\begin{eqnarray*}}
\newcommand{\ee}{\end{eqnarray*}}

\volume{00}

\firstpage{1}

\journalname{Procedia Computer Science}

\runauth{}

\jid{procs}

\jnltitlelogo{Procedia Computer Science}

\usepackage{amssymb}
\usepackage{amsmath}

\usepackage[figuresright]{rotating}

\begin{document}

\begin{frontmatter}



\dochead{}

\title{On angular ordering in medium-induced radiation}


\author{Konrad Tywoniuk\fnref{label2}}
\ead{konrad.tywoniuk@thep.lu.se}
\fntext[label2]{Presently at: Department of Theoretical Physics, Lund University, Solvegatan 14A, SE-223 62 Lund, Sweden.}
\address{Departamento de F\'isica de Part\'iculas, Universidade de Santiago de Compostela,\\ E-15782 Santiago de Compostela, Galicia-Spain}

\begin{abstract}
Medium-induced gluon radiation gives rise to the strong energy loss observed in single hadron spectra in heavy ion collisions. Its angular structure, leading e.g. to intra-jet correlations, is only known for the one-gluon inclusive case. In the vacuum, properties such as angular ordering appear when considering less inclusive distributions due to destructive interference effects. We present a calculation which takes into account interference effects in a medium for a setup in which we consider gluon radiation off a quark-antiquark pair with a given opening angle. Surprisingly, computing the spectrum at leading-log accuracy we find that an ordering exists which is exactly opposite from that in the vacuum. We estimate the magnitude of the effect along with a discussion of the implications of such "anti-angular ordering" on jet physics in heavy ion collisions.
\end{abstract}

\begin{keyword}
QCD \sep jet physics \sep heavy-ion collisions \sep jet quenching

\end{keyword}

\end{frontmatter}


\section{Introduction}
\label{sec:intro}
With the startup of the heavy ion collision program at the Large Hadron Collider (LHC) at CERN nuclear physics has entered a new era where processes involving large momentum transfers are expected to play a central role. One such example is the physics of jet fragmentation which so far have proven to be one of the most accurate tests of perturbative QCD, e.g., in proton-proton or $e^+e^-$ collisions. In these cases, when the fragmentation takes place in vacuum, gluon emissions exhibit soft and collinear divergences which can compensate the smallness of the strong coupling constant and have to be re-summed  \cite{bas83}. Due to interference effects among the emitted gluons the striking feature of strict angular ordering of subsequent emissions arise \cite{Mueller:1981ex,Ermolaev:1981cm}. This restriction on available phase space leads to a strong suppression of soft gluon emissions dubbed the humpbacked plateau. Naively, one would expect a weakening of these effects in the medium due to momentum exchanges and color randomization. We will see, however, that this is not the case and that color interference effects lead to strong modifications of the jet structure. 

On the theory side, efforts to address the question whether and how the jet evolution is altered by the presence of the QGP have been put forward in the past few years. Although different medium effects could lead to changes in the jet properties, the modification of the gluon radiation pattern is expected to be of main relevance. This modification is only known for the one gluon inclusive radiation off a fast quark or gluon \cite{bdmps,zakharov,gyu00,Wiedemann:2000za}. In contrast to the vacuum, both the soft and collinear divergencies of the resulting spectrum are screened due to the gluon interaction with the medium.

Clearly, color coherence effects among the different partons in the cascade are not addressed in this setup since only a single emitter is considered, nor is the presence of an ordering variable for subsequent emissions. As a first step towards a jet calculus in medium, recently we calculated the medium-induced  radiation spectrum off a quark-antiquark ($\qqb$) pair \cite{MehtarTani:2010ma}. In the vacuum, this setup provides a simple laboratory for the intrajet coherent cascade, encompassing, in particular, the key feature of angular ordering \cite{Dokshitzer:1991wu}.

Compared to the state-of-the-art jet measurements in proton--(anti)proton collisions, jet physics in heavy-ion collisions is still in its early stages. Indeed, the investigation of these new possibilities started at RHIC \cite{Putschke:2008wn} and progress rapidly with the advent of the LHC experimental program, with its high capabilities of jet measurements even in high-multiplicity events. 
This motivates a fresh look on possible novel features of medium effects on jets which can provide interesting tools to probe the nature of the quark-gluon plasma (QGP).

\section{Radiation off a $\qqb$ antenna in the presence of a medium}
Let us first briefly review the vacuum emission pattern of gluon emission with  momentum $k\equiv(\omega,\bf k)$, $\omega$ being its energy and $\bf k$ its 3-momentum vector, off a $q \bar q$ pair with momenta $p\equiv(E, \bf p)$ and $\bar p\equiv( \bar E, \bar{\bf p})$, respectively. The gluon spectrum off a $\qqb$ antenna in a singlet state is given by
 \beq
\label{eq:spectrum-vac1}
(2\pi)^2\,\omega\frac{dN^\text{vac}}{d^3k}=\frac{\alpha_s C_F}{\omega^2} \frac{2\, n_q\cdot n_{\bar q}}{(n_q\cdot n)(n_{\bar q}\cdot n)} \,,
\eeq
where $n_q \equiv p/E$, $n_{\bar q}\equiv \bar p/\bar E$ and $n\equiv k/\omega$.  The color factor $C_F=(N_c^2-1)/2N_c$ denotes the color charge of the quark. The cross section in Eq.~(\ref{eq:spectrum-vac1}) exhibits an apparent double collinear singularity.  The two poles can be split into two separate terms which comprise the quark and the antiquark collinear divergences, respectively. Averaging the collinear singular part of Eq.~(\ref{eq:spectrum-vac1}) along the direction of, e.g., the quark over the azimuthal angle leads to gluon emissions restricted to a cone defined by twice the opening angle of the $\qqb$ pair, $\theta_{q\bar q}$. This owes to the fact that large-angle radiation is suppressed since it does not resolve the internal structure of the pair. Thus, the corresponding gluon emission probability off the quark in vacuum reads  
\beq
\label{eq:spectrum-vac2}
dN^{\text{vac}}_q=\frac{\alpha_sC_F}{\pi} \frac{d\omega}{\omega}\frac{\sin\theta \ d \theta}{1-\cos\theta}\  \Theta(\cos\theta-\cos\theta_{q\bar q}),
\eeq
which exhibits a double logarithmic singularity, namely a soft divergence, when $\omega\to 0$, and a collinear divergence, when $\theta\to 0$, where $\theta$ is the angle between the quark and the emitted gluon.

The medium modification of this cross section will be treated in the approximation of small opening angles, $\theta_\qqb\ll 1$, for simplicity. The interaction of the antenna components with the medium is calculated at first order in opacity, i.e., assuming one gluon exchange \cite{MehtarTani:2010ma}. The medium background field is described by ${\cal A}_M(x^+,\q)$, where the $x^+$ is the longitudinal position and $\q$ the exchanged transverse momentum. The medium average is defined as
\beq
\langle {\cal A}^a_\sM(x^+,\q) {\cal A}^{\ast b}_\sM(x'^+,\q')\rangle\equiv \delta^{ab}\,n_0 \,m_D^2\,\delta(x^+-x'^+)\, (2\pi)^2 \,\delta^{(2)}(\q-\q'){\cal V}(\q)\,,
\eeq
where ${\cal V}(\q)=1/(\q^2+m_D^2)^2$ is the medium gluon potential squared, $m_D$ is the Debye mass and $n_0$ is the 1-dimensional density of scattering centers. Then the amplitude of gluon radiation off a quark is given by
\beq
\label{eq:amplitude}
{\cal M}^{a}_{q,(1)} \!&=&\!ig^2 f^{abc} t^c \, \frac{p^+}{k^+} \int  \frac{d^2\q}{(2\pi)^2}\int_{0}^{L^+} \!\! dx^+  {\cal A}^b_\sM(x^+,\q)
\left[ \frac{{\bs \nu}\cdot {\bs \epsilon}_\perp}{p\cdot v} \left(1-e^{i\frac{p\cdot v}{p^+}x^+}\right)+\frac{{\bs \kappa}\cdot {\bs \epsilon}_\perp}{p\cdot k}e^{i\frac{p\cdot v}{p^+}x^+}\right] e^{i (k^-- v^-) x^+} \,, 
\eeq
where $\kappa^i = k^i - (k^+/p^+) \,p^i$ ($i=1,2$), $v \equiv \left(v^+=k^+,v^-= \v^2\big/2k^+, \v=\k-\q\right)$ and $\nu^i =v^i-(k^+/p^+)\,p^i$\footnote{Light-cone components are defined as $p^\pm = (p^0 \pm p^3)/\sqrt{2}$.}. In Eq.~(\ref{eq:amplitude}), $L^+=\sqrt{2}L$, where $L$ is the medium size. The amplitude for gluon radiation off the antiquark  is deduced from $\mathcal{M}_q$ by the substitutions $p \to \bar p$ and an overall change of sign. The first term  in Eq.~(\ref{eq:amplitude}) corresponds to the interaction of the emitted gluon with the medium, denoted ${\cal M }^{g}_q$, while the second term corresponds to the interaction of the quark before the bremsstrahlung emission of the gluon, denoted ${\cal M}_q^{\text{brem}}$.
The contact terms, being the interference between the gluon emission amplitude in vacuum and the one accompanied by two-gluon scattering, are essential for unitarity and simply lead to a redefinition of the potential such that $ {\cal V}(\q)\to{\cal V}(\q)-\delta(\q)\int d^2\q' {\cal V}(\q')$, which guarantees that the spectrum is finite in the $\q \to0$ limit.

Squaring the amplitude and summing over the polarization vector, $|\mathcal{M}|^2=|\mathcal{M}_q|^2 + |\mathcal{M}_{\bar q}|^2 + 2\,\text{Re}\,\mathcal{M}_q \mathcal{M}_{\bar q}^\ast$, firstly we recover the Gyulassy-Levai-Vitev (GLV) spectrum \cite{gyu00,Wiedemann:2000za} for the quark and the antiquark, respectively (two first terms on the r.h.s.). The sum of the two we denote by $\mathcal{I}_{\text{GLV}}$. Additionally, we also get novel contributions stemming from the interference. The latter can be further divided into two contributions, namely $\mathcal{I}_{\text{brems}} = 2\,\text{Re}\,{\cal M}^{\text{brem}}_q {\cal M}^{*\text{brem}}_{\bar q}$, which is the only term exhibiting a soft divergence, and the remaining ones, involving at least one gluon interaction with the medium, denoted by $\mathcal{I}_{\text{interf}}$.

\begin{figure}
\centering \includegraphics[width=0.56\textwidth]{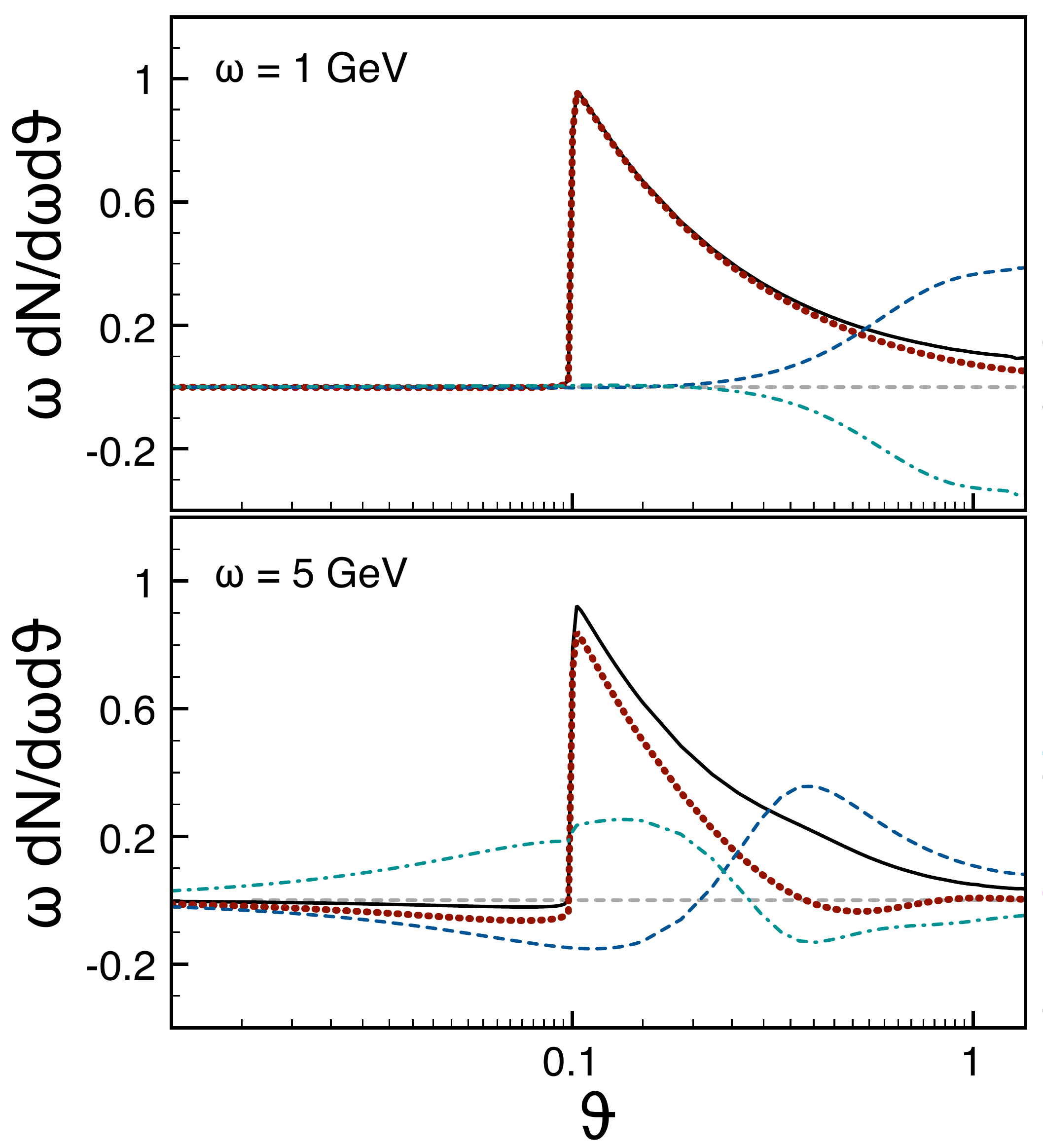}
\caption{The angular distribution of the medium-induced gluon spectrum for $\omega$=1 and 5 GeV for a $q\bar q$ pair with opening angle $\theta_{q\bar q}= 0.1$, see the text for details. The dotted (red) line corresponds to the dominant contribution in the soft-limit, $\mathcal{I}_{\text{brems}}$, see Eq. (\ref{eq:SpectrumMedSoft}), while the short-dashed (blue) curve is the sum of GLV contributions from the quark and the antiquark, $\mathcal{I}_{\text{GLV}}$, and the dash-dotted (green) curve depicts the remaining terms, $\mathcal{I}_{\text{interf}}$. The solid line corresponds to the total spectrum. Figure taken from \cite{MehtarTani:2010ma}.}
\label{fig1}
\end{figure}

The three contributions are plotted in Fig. \ref{fig1}, where we have evaluated the angular distribution of the full spectrum off a $q \bar q$ pair with opening angle $\theta_{q\bar q} = 0.1$ traversing a medium with thickness $L = 4$ fm ($m_D=0.5$ GeV, $\alpha_s = 1/3$ and $n_0L=1$) numerically for two gluon energies. We note that, in both cases, the three terms add up to zero at small angles, leaving the cone delimited by the pair angle empty. The distribution jumps from zero inside the cone to a maximum value at $\theta=\theta_{q\bar q}$, it then drops as $ 1/\theta$ for $\theta>\theta_{q\bar q}$. This vacuum-like pattern persists at the higher energy, see Fig.~\ref{fig1} bottom, caused by an intricate cancellation between the different contributions which differs notably from the single-particle GLV spectrum. Specifically, there is no medium broadening of the jet in this limit.

In fact, these general features can be understood in the soft limit, i.e., $\omega\to 0$, where the dominating contribution to the spectrum is simply given by $\mathcal{I}_{\text{brems}}$, namely
\beq
\label{eq:SpectrumMedSoft}
\omega\frac{dN^{\text{med}}}{d^3k}= \frac{8\pi C_AC_F\, \alpha_s^2 \,n_0\, m_D^2}{(2\pi)^2 \,(k^+)^2} \, \frac{p^+\bar p^+\, \left(\bkappa \cdot \bbkappa\right)}{(p\cdot k)( \bar p\cdot k)} \int_0^{L^+}\!\!\!\! d x^+ \cos \Omega_0\,x^+ \int \frac{d^2\q }{(2\pi)^2} {\cal V}(\q) \left(1-\cos \Delta\Omega\, x^+ \right) \,,
\eeq
where we have written the contact term explicitely, and where $\Omega_0= p\cdot k/p^+ - \bar p \cdot k/\bar p^+$, $\Delta\Omega= \p\cdot \q/p^+ - \pb \cdot \q/\bar p^+$.
The soft divergence in Eq.~(\ref{eq:SpectrumMedSoft}) is manifest. Note that in the soft limit $\Omega_0 \to 0$  and the integrals in Eq.~(\ref{eq:SpectrumMedSoft}) are straightforward, yiedling $ L^{+} r_\perp^2 [\ln \left(1/r_\perp m_D\right)+\text{const.}]/24\pi$,
where we have assumed the quark momentum to be along the $z$ axis for simplicity, i.e., $p_\perp=p^-=0$. Let us finally turn to the angular structure of Eq.~(\ref{eq:SpectrumMedSoft}), second term on the r.h.s. Dividing the term symmetrically into a quark and an antiquark contribution and afterwards averaging over the azimuthal angle along the direction of, e.g., the quark, we obtain in the small angle limit ($\theta_{q\bar q} \ll 1$ and $\theta \ll 1$)
\beq
\label{eq:AngularMed}
\left\langle \frac{1}{2}\frac{p^+\bar p^+ \, (\bkappa \cdot \bbkappa)}{(p\cdot k)(\bar p\cdot k)} \right\rangle_\varphi \simeq  \frac{\Theta(\cos\theta_{q\bar q}-\cos\theta)}{1-\cos\theta} \;,
\eeq
thus recovering a structure very similar to the vacuum one, cf. Fig.~\ref{fig1}. However, the medium-induced soft gluon radiation off the quark is suppressed inside the cone of opening angle $\theta_{q\bar q}$, as opposed to the standard angular structure obtained in vacuum, see Eq.~(\ref{eq:spectrum-vac2}). Furthermore, due to this feature, the collinear pole in Eq.~(\ref{eq:AngularMed}) is automatically cut off. 
Thus, when $\omega \rightarrow 0$ the medium-induced gluon emission off the quark can be written as \cite{MehtarTani:2010ma}
\beq
\label{eq:nqmed}
dN^{\text{med}}_q=\frac{\alpha_sC_F }{\pi}A^{\text{med}}\frac{d\omega}{\omega}\frac{\sin\theta \ d \theta}{1-\cos\theta} \ \Theta(\cos\theta_{q\bar q}-\cos\theta)
\eeq
where $A^{\text{med}} = \alpha_sC_A  n_0 m_D^2 L^{+} r_\perp^2 [\ln \left(1 \big/ (r_\perp m_D\right)+\text{const.}] /6$ is the forward dipole scattering amplitude in the adjoint representation for a dipole of size $r_\perp = \theta_{q\bar q} L$.

\section{Discussion}
\label{sec:discussion}

The full gluon spectrum in the presence of a medium is thus given by $dN_q^{\text{tot}} = dN_q^{\text{vac}} + dN_q^{\text{med}}$. The medium-induced spectrum off the antenna, obtained in \cite{MehtarTani:2010ma}, thus establish that there is a {\it strict geometrical separation} between vacuum and medium-induced radiation, cf. Eq.~(\ref{eq:spectrum-vac2}), and that there arises a soft divergence in the latter. Here, it is also worth noting that $dN_q^{\text{med}}$ is proportional to the color factor of the emitter, in this case that of a quark, $C_F$. For a sufficiently opaque medium, the full spectrum will become increasingly decoherent, see also \cite{Leonidov:2010}.

Although the calculation above \cite{MehtarTani:2010ma} has been done for a $\qqb$ pair in a singlet state, it is easy to show that in the soft limit for the octet case, where we can neglect radiation off the initial gluon due to its large virtuality, the same conclusions apply, the main difference being related to the color algebra. Therefore we expect this effect to be of great relevance for jet fragmentation in heavy-ion collisions, leading, e.g., to a distortion of the soft part of the intrajet spectrum.

\subsection*{Acknowledgements}
K.~T. would like to thank N. Armesto, A. Kovner and U.~A. Wiedemann for illuminating discussion. Financial support from the organizers of HP2010 is greatly appreciated.




\bibliographystyle{elsarticle-num}
\bibliography{<your-bib-database>}



\end{document}